\documentclass[12pt,preprint]{aastex}

\begin{document}

\title{A Geometric Determination of the Distance to the Galactic Center
\footnote{based on observations obtained at the Very Large Telescope (VLT) of
the European Southern Observatory, Chile}
}

\author{
F.Eisenhauer, R.Sch\"{o}del, R.Genzel\altaffilmark{2}, T.Ott, M.Tecza and R.Abuter
}
\affil{
Max-Planck Institut f\"{u}r extraterrestrische Physik (MPE), Garching, Germany
}
\author{
A.Eckart
}
\affil{
1.Physikalisches Institut der Universit\"{a}t K\"{o}ln, Germany
}
\author{
T.Alexander
}
\affil{
Faculty of Physics, Weizmann Institute of Science, Rehovot, Israel
}

\altaffiltext{2}{also Department of Physics, University of California, Berkeley (USA)}

\begin{abstract}

We report new astrometric and spectroscopic observations of the star S2
orbiting the massive black hole in the Galactic Center, which were taken at
the ESO VLT with the adaptive optics assisted, near-IR camera NAOS/CONICA
and the near-IR integral field spectrometer SPIFFI. We use these data to
determine all orbital parameters of the star with high precision, including
the Sun-Galactic Center distance, which is a key parameter for calibrating
stellar standard candles and an important rung in the extragalactic distance
ladder. Our deduced value of $R_{o}$=8.0$\pm $0.4 kpc is the most accurate
primary distance measurement to the center of the Milky Way and has minimal
systematic uncertainties of astrophysical origin. It is in excellent
agreement with other recent determinations of $R_{o}$.
\end{abstract}


\section{Introduction}

The distance between the Sun and the Galactic Center ($R_{o}$) is a
fundamental parameter for determining the structure of the Milky Way.
Through its impact on the calibration of the basic parameters of standard
candles, such as RR Lyrae stars, Cepheids and giants the Galactic Center
distance also holds an important role in establishing the extragalactic
distance scale. Ten years ago Reid (1993) summarized the state of our
knowledge on $R_{o}$. At that time the only primary (geometric) distance
indicator to the Galactic Center came from the 'expanding cluster parallax'
method applied to the H$_{2}$O masers in SgrB2, believed to lie within 
$\sim$0.3 kpc of the Galactic Center (G\"{u}sten \& Downes 1980). Reid
et al. (1988a, b) determined values of 7.1 and 6.5 kpc for the distances to
the masers in SgrB2(N) and SgrB2(M), respectively, with a combined
statistical and systematic (1$\sigma $) uncertainty of $\pm $1.5 kpc. In
addition there existed a number of secondary (standard candle)
determinations, based on RR-Lyrae stars, Cepheids, globular clusters and
giants, as well as a number of tertiary indicators, derived from theoretical
constraints (e.g. Eddington luminosity of X-ray sources, Galaxy structure
models). From all these measurements Reid inferred a best overall estimate
of 8.0 kpc, with a combined uncertainty of $\pm $0.5 kpc. In the time since
1993 Genzel et al. (2000) reported a primary (statistical parallax)
distance, $R_{o}$=8.0$\pm $0.9 kpc (statistical error bar), based on a
statistical comparison of proper motions and line-of-sight velocities of
stars in the central 0.5 pc of the Galaxy. Carney et al. (1995) and McNamara
et al. (2000) found secondary distances of 7.8 and 7.9 kpc ($\pm $0.7 kpc)
from RR Lyrae and $\delta $-Scuti stars. Feast \& Whitelock (1997) used
Cepheids and a Galactic rotation model, updated for the new Hipparcos local
distance scale, to obtain $R_{o}$=8.5$\pm $0.5 kpc. Paczynski \& Stanek
(1998) and Stanek \& Garnavich (1998) combined measurements of red clump
stars with the Hipparcos scale to obtain $R_{o}$=8.2 kpc, with a claimed
combined statistical and systematic uncertainty of $\pm $0.21 kpc.

We report here the first primary distance measurement to the Galactic Center
with an uncertainty of only 5\%. This determination has become possible
through the advent of precision measurements of proper motions and
line-of-sight velocities of the star S2. This star is orbiting the massive
black hole and compact radio source SgrA$^{\ast }$ that is located precisely
at the center of the Milky Way. As discussed by Salim \& Gould (1999), the
classical 'orbiting binary' technique can then be applied to obtain an
accurate determination of $R_{o}$ that is essentially free of systematic
uncertainties in the astrophysical modeling. The essence of the method is
that the star's line-of-sight motion is measured via the Doppler shift of
its spectral features in terms of an absolute velocity, whereas its proper
motion is measured in terms of an angular velocity. The orbital solution
ties the angular and absolute velocities, thereby yielding the distance to
the binary. Sch\"{o}del et al. (2002) found that S2 is on a highly
elliptical Kepler orbit around SgrA$^{\ast }$, with an orbital period of
about 15 years and a peri-center distance of about 17 light hours. Ghez et
al. (2003) confirmed and improved the Sch\"{o}del et al. (2002) results and
reported the first spectroscopic identification and line-of-sight velocity
measurement of S2. S2 appears to be a 15-20 $M_{\odot }$ main sequence O8-B0
star whose line-of-sight velocity can be inferred in a straightforward
manner from the HI Br$\gamma $ absorption in the star's K-band (2.2$\mu $m)
spectrum. With additional line-of-sight velocity and proper motion data it
has now becomes feasible to make a precision estimate of all orbital
parameters, including the distance to S2/SgrA$^{\ast }$.

\section{Observations}

\subsection{NAOS/CONICA adaptive optics imaging}

We observed the Galactic Center on Yepun (UT4 of the VLT) with the
NAOS/CONICA (NACO, Lenzen et al. 1998, Rousset et al. 1998) near-infrared,
adaptive optics imager on March 17/18 2003 (2003.21) and May 8/9 2003
(2003.35). We observed through the H-band (1.65$\mu $m) filter and used the
infrared wavefront sensor to correct in K-band\ on the bright supergiant
IRS7, located $\sim$5.6'' north of S2/SgrA$^{\ast }$. In both runs the
seeing was $\leq $0.5'', resulting in H-band Strehl ratios between 0.3 and
0.5. After producing final maps with the shift-and-add technique, we
deconvolved the images with a linear Wiener filter and a Lucy-Richardson
algorithm. We extracted stellar positions from the deconvolved images with
the program STARFINDER (Diolaiti et al. 1995). We also applied this
technique to all 2002 NACO Galactic Center imaging data described by Sch\"{o}%
del et al. (2003), thereby improving slightly their results. All positions
prior to 2002 are from the observations with the SHARP instrument on the ESO
3.5m NTT, as reported by Sch\"{o}del et al. (2003). Our positional errors
are a combination of fit errors and errors in placing S2 in a common
infrared astrometric frame, resulting in overall errors for NACO of 1-3 mas
(1$\sigma $) in 2003 and 3-7 mas in 2002 (when the performance of the then
new NACO instrument was still not optimized), and 6-10 mas for the 1992-2001
SHARP/NTT measurements. In addition, there is a $\pm $10 mas absolute
uncertainty between the infrared and radio astrometric frames (Reid et al.
2003a). Figure 1 is a plot of the positions of S2 between 1992 and 2003.

\subsection{SPIFFI integral field spectroscopy}

On April 8/9 2003 (2003.27) we observed the Galactic Center on Kueyen (UT2
of the VLT) with the new MPE integral field spectrometer SPIFFI (Thatte et
al. 1998, Eisenhauer et al. 2000). Briefly, SPIFFI uses a reflective image
slicer and a grating spectrometer to simultaneously obtain spectra with 1024
spectral elements for a contiguous 32$\times$32 pixel, two dimensional field on
the sky. In very good seeing conditions (0.4''-0.5'' in the optical) we
observed with a pixel scale of 0.1'', resulting in a 2$\mu $m FWHM of
0.25''-0.3''. The FWHM spectral resolution was 85 km/s, sampled at 34 km/s.
We dithered about two dozen exposures of 1 minute integration time each to
construct a mosaicked data cube of the central $\sim$8''. Data
reduction used the new SPIFFI analysis pipeline, as well as IRAF tools. We
determined the wavelength calibration from arc lamps and OH sky emission
lines, resulting in an accuracy of $\pm $7 km/s. The effective integration
time toward the central part of the mosaic near SgrA$^{\ast }$ was about 15
minutes. We used both the flat spectrum star IRS16CC and the ATRAN
atmospheric code to correct for atmospheric absorption. We then extracted
spectra toward S2 and several 'off' positions nearby in apertures of about
0.2'' in diameter. We analyzed the direct as well as the off-subtracted
spectra. Off-subtraction is important for removing the extended nebular Br$%
\gamma $ emission from the SgrA West mini-spiral. For our spring 2003 data,
direct and off-subtracted spectra were identical near the star's Br$\gamma $
absorption line at v$_{LSR}$= -1545$\pm $25 km/s, far off the nebular
contamination. The direct SPIFFI spectrum is shown in the left inset of
Figure 2.

\subsection{NACO long slit spectroscopy}

We took K-band grism spectroscopy of S2 with NACO on Yepun on May 8/9 2003
(2003.35). As for the H-band imaging, the optical seeing was $\sim$%
0.4''-0.5'' and we used the infrared wavefront sensor on IRS7. We chose the 86
mas slit, resulting in 210 km/s resolution sampled at 69 km/s per pixel. The
spatial pixel scale was 27 mas and we placed the slit at position angle 78$%
^{o}$ through S2. We integrated for about 5 minutes per readout and nodded
the slit by $\pm $5''. We accumulated 30 minutes of on-source integration.
Data analysis employed standard IRAF longslit tools. We calibrated on arc
lamps, resulting in a velocity accuracy of $\pm $10 km/s. We corrected for
atmospheric absorption by dividing the Galactic Genter spectra by an early
type star observed at the same airmass. The inferred LSR velocity of the Br$%
\gamma $ absorption of S2 was -1512$\pm $35 km/s. Again, no correction for
nebular emission was necessary. The nodded NACO spectrum in a 0.086''$\times$%
0.1'' aperture is shown in the right inset of Figure 2.

\section{Results}

\subsection{Geometric Distance Estimate to the star S2}

For the analysis of our measurements we fitted the positional and
line-of-sight velocity data to a Kepler orbit, including the Galactic Center
distance as an additional fit parameter. In principle the dynamical problem
of two masses orbiting each other requires the determination of 14
parameters: 6 phase space coordinates for each mass, plus the values of the
two masses (see Salim \& Gould 1999). At the present level of accuracy, four
of these 14 parameters can be safely neglected: the mass of the star (since m%
$_{S2}$/M$_{SgrA^{\ast }}$ $\sim$5$\times$10$^{-6}$) and the three velocity
components of SgrA$^{\ast }$. Radio interferometry of SgrA$^{\ast }$ with
respect to background quasars has established that after subtraction of the
motions of Earth and Sun around the Galactic Center the proper motion of SgrA%
$^{\ast }$ is $\leq $ 20 km/s in the plane of the Galaxy and $\leq $8 km/s
toward the Galactic Pole (Backer \& Sramek 1999, Reid et al. 1999, Reid et
al. 2003b). This implies v$_{SgrA^{\ast }}$ $\sim $10$^{-2}$ v$_{S2}$.
Likewise the uncertainty in the local standard of rest velocity ($\leq $10
km/s) can also be neglected at the present level of analysis (see Salim \&
Gould 1999). In the actual orbital fitting we solve for the geometric
parameters of the orbit (Table 1), as well as the time of peri-center
passage. The central mass is a dependent variable that is calculated from
the semi-major axis and period, using Kepler's 3$^{rd}$ law. Our measurement
constraints consist of the 18 (x2) S2 positions between 1992 and 2003 and 4
line-of-sight velocities: 2 from Ghez et al. (2003), 1 from SPIFFI and 1
from NACO. This leaves us to fit 10 parameters with 40 data points,
resulting in an over-constrained problem with 30 degrees of freedom. For
fitting the orbit of S2, we expanded our existing IDL program
codes for Keplerian orbits (see Sch\"{o}del et al. 2003) to take into
account the additional information provided by line-of-sight velocity
measurements. The errors of the orbital parameters are based on an analysis
of the covariance matrix. The uncertainty in $R_{o}$ is most strongly
correlated with the semi-major axis $a$, the inclination $i$, and the time
of peri-center approach $T_{peri}$, all of which are well determined in our
model. Table 1 is a list of the fitted parameters of the S2 orbit, the mass
and location of the central object and the distance to the Galactic Center.
Figures 1 and 3 show the best-fit orbital and line-of-sight velocity curves
derived from the fit parameters in Table 1, superposed on our data. The
accuracy of the orbital parameters in Table 1 is 3 to 6 times better than
those in Sch\"{o}del et al. (2002), and 1.3 to 2 times better than those in
Ghez et al. (2003).

\subsection{Statistical Parallax to the Central Star Cluster}

In addition to the $R_{o}$ value from the S2 orbit, we also report an update
of the statistical parallax distance to the stars in the central 0.5 pc. We
used the proper motion and line-of-sight velocity data base of Ott et al.
(2003) and complemented these by additional $\sim$100 velocities of
early type and late type stars in the central 10'' extracted from the new
SPIFFI data cube discussed above (Abuter et al. 2003). We now have 106 late
type stars and 27 early type stars with all three velocities. For these two
sets we can apply the anisotropy independent distance estimator introduced
by Genzel et al. (2000),

\ \ \ \ \ \ \ \ \ \ \ \ \ \ \ \ \ \ \ \ \ \ \ \ \ \ \ \ \ \ \ \ $\left( 
\frac{R_{o}}{8kpc}\right) $\ \ $=$ $\left( \frac{<pv_{z}^{2}>_{8}}{%
1/3<pv_{R}^{2}>_{8}+2/3<pv_{T}^{2}>_{8}}\right) ^{0.5}$\ \ \ \ \ \ \ \ \ (1),

where $p$ is the sky projected distance of a star from SgrA$^{\ast }$, and $%
v_{z}$, $v_{R}$ and $v_{T}$ are the line- of-sight, sky projected radial,
and sky projected tangential velocities (from proper motions). The symbol $%
<>_{8}$ denotes the ensemble average for an assumed distance of 8 kpc.
Within the radius of influence of the black hole ($\sim$0.5 pc) this
estimator gives an equal weight to stars at different distances from SgrA$%
^{\ast }$. Applying this estimator to the 106 late type stars ($p\leq$10'')
with three velocities yields $R_{o}$=7.1$\pm $0.7 kpc. For the 27 early type
stars we find $R_{o}$=8.0$\pm $1.6 kpc, where the error bars are statistical
in both cases. The early and late type stars have very different dynamical
properties and thus need to be treated separately (Genzel et al. 2003). We
estimate that both values have an additional systematic uncertainty (due to
phase space clumping, possible streaming motions etc.) of $\pm $0.6 kpc. The
statistical parallax method thus gives $R_{o}$=7.2 kpc (with a combined
uncertainty of $\pm $0.9 kpc), in good agreement with the more accurate S2
orbit determination.

\section{Discussion}

The value of $R_{o}$ deduced from the orbit of S2 is 8.0$\pm $0.4 kpc. Our
determination rests on the analysis of a simple dynamic system. The massive
black hole candidate SgrA$^{\ast }$ is located at the very center of the
Milky Way. The distance value and its uncertainty come from a global fit to
all data that includes all parameter interdependencies through the
covariance matrix. Hence we are confident that the deduced errors contain
all sources of uncertainty. The derived distance has no sizeable additional
systematic uncertainties due to the astrophysical modeling. For instance,
deviations of the gravitational potential from that of a point source are
small. Sch\"{o}del et al. (2002, 2003) and Genzel et al. (2003) conclude
that any distributed mass with a density distribution similar to that of the
central stellar cusp cannot contribute more than a few hundred solar masses
within the peri-center distance of S2, or 10$^{-4}$ of the mass of the
central black hole. It is also unlikely that the central mass is a binary
black hole of approximately equal masses, since such binary hole would have
to have a separation less than 10 light hours (constrained by the data) and
would coalesce by gravitational radiation in a few hundred years. The
fractional uncertainty in the value of $R_{o}$ is under 5\%, thus delivering
the most accurate, primary Galactic Center distance measurement so far. It
is gratifying to see how well our value agrees with all other, primary and
secondary distance measurements. This gives confidence in the quality and
robustness of the standard candles methods (RR Lyrae stars, Cepheids, red
clump stars etc.) that are at the key of the second rung of the
extragalactic distance ladder.

Future improvements in the accuracy of $R_{o}$ from the orbit of S2 alone
will be relatively slow. This is because we have already sampled three
quarters of the entire orbit, and have also observed the largest swing in
the line-of-sight velocity curve. As discussed in Salim \& Gould (1999),
further significant improvements to the level of under a few percent can be
expected from combinations of several orbits, since then the number of
degrees of freedom is rapidly increasing (as four fit parameters are common
to all orbiting stars).

\acknowledgments
Acknowledgements. We are grateful to N.Thatte, C.Iserlohe,
J.Schreiber, M.Horrobin, C.R\"{o}hrle, S.Huber and S.Weisz whose work on
SPIFFI was essential for the Galactic center data taken here. We also thank
the Paranal staff and our colleagues from NAOS (D.Rouan, F.Lacombe) and
CONICA (R.Lenzen, P.Hartung) for their support. We thank M.Reid for valuable
comments on this paper, and B.Schutz for comments on the coalescence time
for binary black holes. TA is supported by GIF grant 2044/01, Minerva grant
8484 and a New Faculty grant by Sir H. Djangoly, CBE, London, UK.

\clearpage

\begin{deluxetable}{cc}
\tablecaption{Best Kepler orbit fit to the NACO, SPIFFI and NIRC2 data of S2}
\startdata
degrees of freedom of fit & 30\\
$\chi ^{2}$ & 17.22 \\
$\chi _{red}^{2}$ & 0.58 \\
semi-major axis $a$ & 0.1200 $\pm $ 0.0026 arcsec\\
eccentricity $e$ & 0.880 $\pm $\ 0.006\\
orbital period $P$ & 15.559 $\pm $\ 0.337 yr\\
time of peri-center approach $T_{peri}$ & 2002.329 $\pm $\ 0.011 yr\\
inclination $i$ & -47.9 $\pm $\ 1.3 deg\\
angle of line of nodes $\Omega $ & 45.3 $\pm $\ 1.5 deg\\
angle of nodes to peri-center $\omega $ & 245.1 $\pm $\ 1.6 deg\\
$x_{o}$ & 0.0022 $\pm $\ 0.0012 arcsec\\
$y_{o}$ & -0.0032 $\pm \ $0.0011 arcsec\\
$R_{o}$ & 7.99 $\pm \ $0.38 kpc\\
$M_{o}$ & 3.65 $\pm $\ 0.25 $\times$10$^{6}$ $M_{\odot }$ \tablenotemark{a} \\
\enddata
\tablenotetext{a}{
$M_{o}$ is not a fit parameter but equals $4\pi ^{2}a^{3}/(GP^{2})$ (3$
^{rd}$ Kepler law).}
\end{deluxetable}

\clearpage

\begin{figure}
\epsscale{0.8}
\plotone{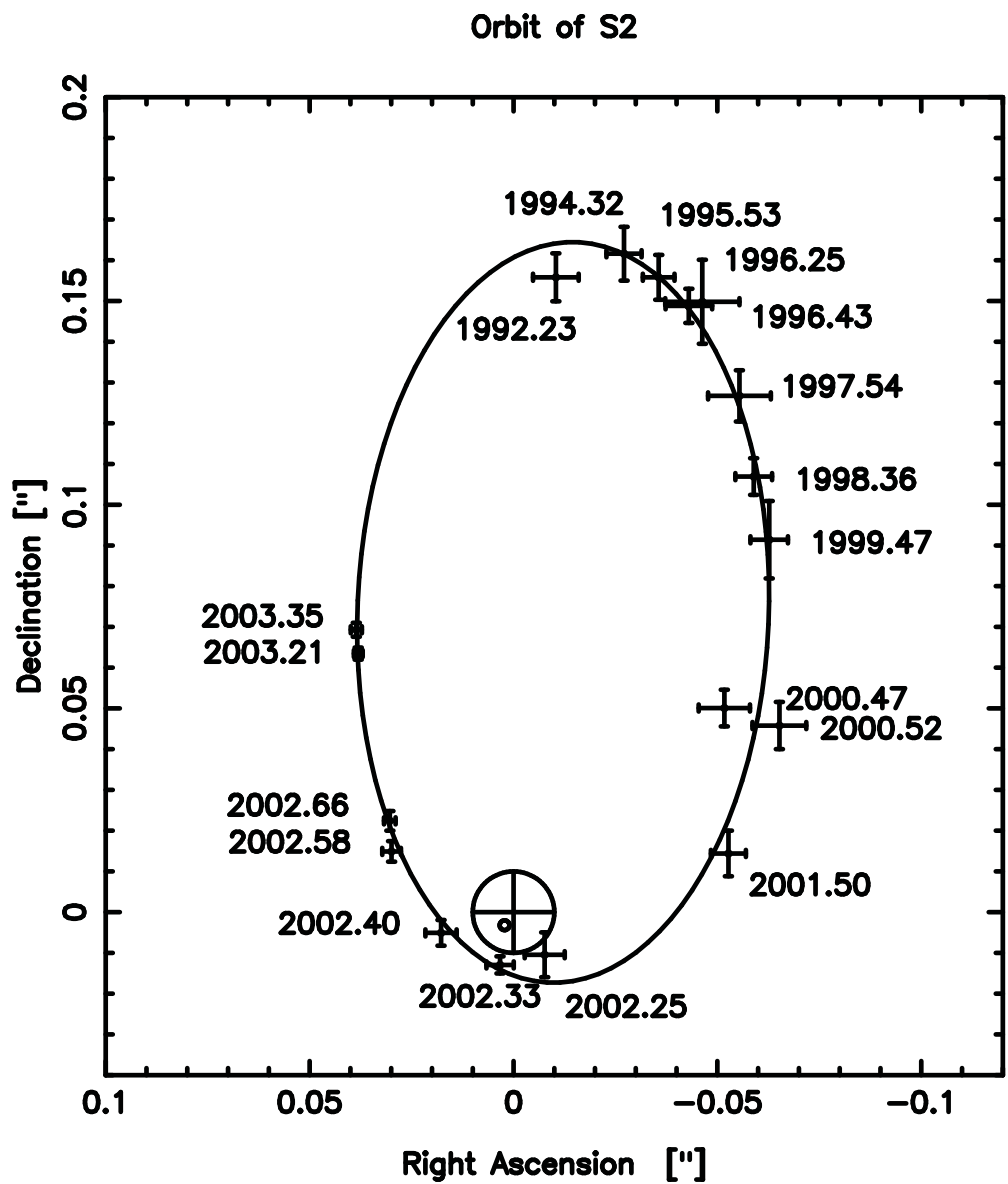}
\caption{
Position measurements of S2 in the infrared astrometric frame.
Crosses (denoting 1$\sigma $ error bars) with dates mark the different position
measurements of S2, taken with the MPE speckle camera SHARP on the 3.5m ESO
NTT (between 1992 and 2001), and with NACO on the VLT (in 2002 and 2003).
The continuous curve shows the best-fit Kepler orbit from Table 1, whose
focus is marked as a small error circle. The focus of the ellipse is within
a few mas at the position of the compact radio source, which is marked by a
large circled cross. The size of the cross denotes the $\pm $10 mas
positional uncertainty of the infrared relative to the radio astrometric
reference frame.}
\end{figure}

\clearpage

\begin{figure}
\plotone{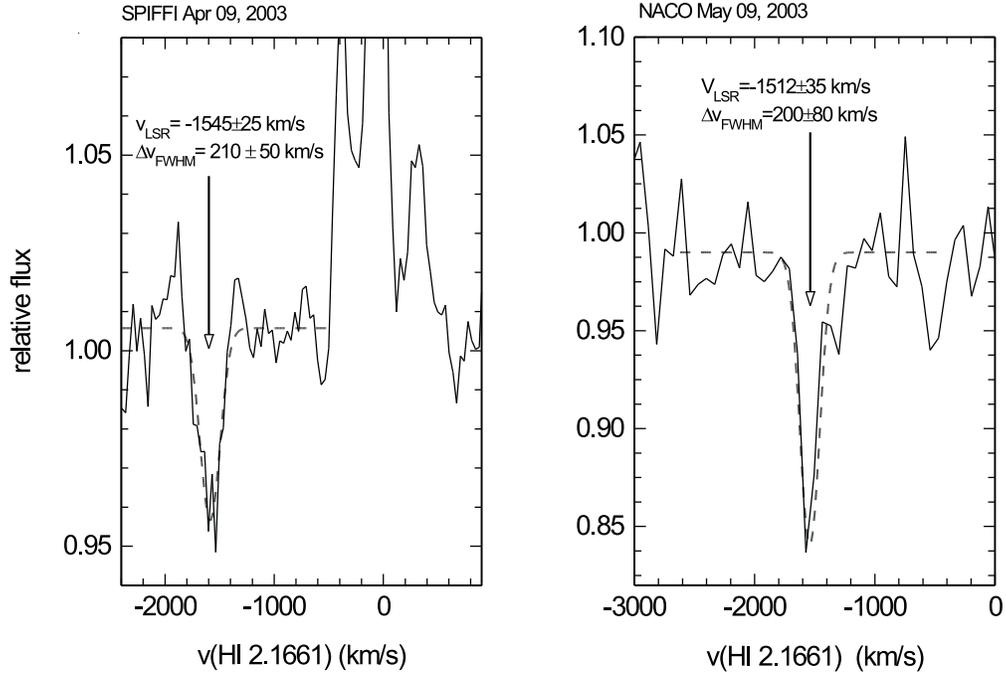}
\caption{
HI Br$\gamma $ absorption spectra of S2, obtained on April 8/9 2003
with the SPIFFI integral field spectrometer on Kueyen (left inset),
and on May 8/9 2003 with the NACO grism on Yepun (right inset). The
SPIFFI spectrum is not off-subtracted and in a 0.2''$\times$0.2'' aperture,
while the NACO spectrum is sky-nodded and in a 0.086''$\times$0.1''
aperture. These differences account for the fact that the mini-spiral
emission features between -400 and +400 km/s LSR are visible in the
SPIFFI data but not in the NACO spectrum. Likewise, dilution of the S2
flux by other nearby sources in the larger SPIFFI beam plausibly
accounts for the shallower absorption relative to NACO.}
\end{figure}

\clearpage

\begin{figure}
\plotone{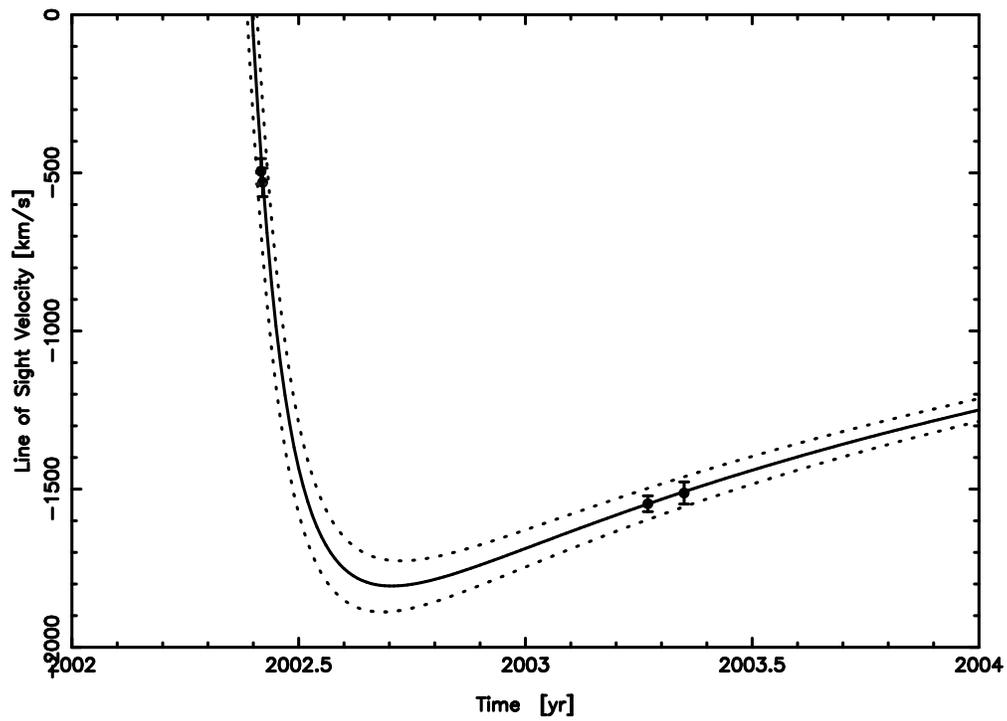}
\caption{
Line-of-sight velocity as a function of time for the best-fit
model of Table 1 (continuous curve), along with the 1$\sigma $ uncertainties
(dotted). Filled circles (with 1$\sigma $ error bars) denote the Keck and
VLT line-of-sight velocities of S2 in 2002 and 2003.}
\end{figure}

\end{document}